\documentclass[aps,prl,twocolumn,superscriptaddress,nofootinbib,showpacs,showkeys,preprintnumbers]{revtex4-1}
\usepackage{graphicx}  
\usepackage{dcolumn}   
\usepackage{bm}        
\usepackage{amsmath, amsthm, amssymb}
\usepackage{lineno}
\usepackage{draftcopy}

\def\lsim{\mathrel{\rlap{
\lower4pt\hbox{\hskip-3pt$\sim$}}
\raise1pt\hbox{$<$}}}     
\def\gsim{\mathrel{\rlap{
\lower4pt\hbox{\hskip-3pt$\sim$}}
\raise1pt\hbox{$>$}}}     

\usepackage{color}
\definecolor{dgreen}{cmyk}{1.,0.,1.,0.2}        
\definecolor{orange}{cmyk}{0.,0.353,1.,0.}    

\usepackage[bookmarks]{hyperref}

\begin{document}
\setlength{\linenumbersep}{5pt}

\title{Charge-dependent directed flow in asymmetric nuclear collisions}

\author{V. Voronyuk}
\affiliation{ Joint Institute for Nuclear Research,  Dubna,
Russia}
\affiliation{Bogolyubov Institute for Theoretical Physics, Kiev,
Ukraine}

\author{V. D. Toneev}
\affiliation{
Joint Institute for Nuclear Research,
RU-141980 Dubna, Moscow Region, Russia
}

\author{S. A. Voloshin}
\affiliation{Wayne State University, Detroit, Michigan, USA  }

\author{W. Cassing}
\affiliation{Institute for Theoretical Physics, University of
Giessen, Giessen, Germany}

\begin{abstract}
The directed flow of identified hadrons is studied within the
parton-hadron-string-dynamics (PHSD) approach for the asymmetric system Cu+Au
in non-central  collisions at $\sqrt{s_{NN}}$ = 200 GeV.
It is emphasized that due to the difference in the number of protons of the colliding
nuclei an  electric field emerges which is directed from
the heavy to the light nucleus. This strong electric field is only present for about 0.25 fm/c at $\sqrt{s_{NN}}$ = 200 GeV and leads to a splitting
of the directed flow $v_1$ for particles with the same mass but opposite
electric charges in case of an early presence of charged quarks and antiquarks.
The microscopic calculations of the directed flow for
$\pi^\pm, K^\pm, p$ and $\bar{p}$ are carried out in the PHSD by taking
into account the  electromagnetic field induced by the
spectators as well as its influence on the hadronic and partonic quasiparticle trajectories.
It is shown that the splitting of the directed flow as a function of
pseudorapidity $\eta$ and in particular as a function of the transverse momentum $p_t$ provides
a direct access to the electromagnetic response of the very early (nonequilibrium) phase of relativistic heavy-ion collisions and allows to shed light on the presence (and number) of electric charges in this phase.
\end{abstract}

\maketitle

\section{Introduction}

An important tool to probe the hot, dense matter created in
heavy-ion collisions is the study of the particle azimuthal
angular distribution in momentum space with respect to the
reaction plane~\cite{VPS10,So10}. In high energy
nuclear collisions  the energy density
reaches  values above 1 GeV/fm$^3$ at which the quantum
chromodynamics (QCD) -- the theory of the strong interactions --
predicts a phase transition from normal hadronic matter to a
deconfined state of  quarks and gluons, the
quark-gluon plasma (QGP). By now, a large amount of experimental
data has been obtained at the relativistic-heavy-ion collider (RHIC)
and the large-hadron-collider (LHC) on the properties of strongly heated and highly
compressed matter what allows  to extract information on the
equation of state of the excited matter as well as on the transport properties
of the partonic degrees of freedom. However, the detailed
properties of this new phase are still  far from being fully understood.

A large elliptic flow $v_2$ of charged hadrons -- observed at RHIC --
testifies the  collective nature of the strong
interaction at  high energies. As the analysis shows~\cite{VPS10,So10},
the excited matter behaves like a colored almost perfect
liquid~\cite{PC05} (the so called strongly coupled QGP -- sQGP) rather
than a weakly interacting parton gas. The elliptic flow has been
measured by many collaborations at energies from the alternating
gradient synchrotron (AGS) to the LHC and
the scaling properties of flow harmonics, their dependence on
centrality, rapidity and particle species have been studied.
Higher harmonics, $v_n (n>2)$, have been also intensively
explored especially in recent years, when it became clear that odd harmonics
are sensitive to fluctuations in the initial conditions.
It finally turns out that the study of azimuthal asymmetries is closely
related to structures in two-particle
correlations of hard and semi-hard
processes  in the partonic phase~\cite{So10}.

In  more recent times much attention has been paid again to the directed
flow \cite{RR97,HWW99} of identified particles  and the precise STAR
measurements performed within the Beam Energy Scan (BES) program in
the energy range $\sqrt{s_{NN}}=$7.7-39 GeV ~\cite{v1STAR} have received
a large attention. The directed flow refers to a
collective sidewards deflection of particles and is characterized by
the first-order harmonic $v_1$ of the Fourier expansion of the
particle azimuthal angular distribution with respect to the reaction
plane. An analysis~\cite{v1KCT} of these data within the microscopic PHSD
transport approach and the collective three-fluid dynamics (3FD) model
reproduces the general trend in the differential $v_1(y)$ excitation
function and leads to an almost quantitative agreement especially at
higher energies where the partonic phase is dominant.
We recall that 3FD hydrodynamics \cite{Yuri} shows a high
sensitivity to the nuclear equation of state (EoS) and
provides the best results employing a
crossover for the quark-hadron phase transition in the
model. Note also that a crossover transition is implemented by default
in the PHSD approach. This flow analysis has shown no indication of
a first order phase transition~\cite{v1KCT} as anticipated long
before~\cite{Rischke,HS05}. In all cases mentioned above only
symmetric nuclear collisions have been considered. The most important
parameters being varied in these analyses are the collision energy,
the size of the system, and the centrality of the collision.

An additional insight into the mechanism of particle production in
relativistic heavy-ion collisions and the electromagnetic response
for very early times can be gained from interactions of
nuclei with different sizes, e.g. Cu+Au collisions at the top RHIC
energy of $\sqrt{s_{NN}}=$200
GeV. The study of the charged hadron multiplicity
distributions and correlations in the longitudinal direction for
asymmetric collisions gives additional constraints to
the mechanism of the energy deposition in the early stage of the
reaction~\cite{CK06,Bo12,STV14,Io13} by exploiting the particle
asymmetry in pseudorapidity. Also, global characteristics
of the directed flow are of substantial interest~\cite{TY11,LO11,HNM11}.
But it is more important, as expected in Refs.\cite{BW10,ARG10}, that the
directed flow in asymmetric collisions may be partly generated by a specific
source. An electric field -- arising from the difference in charges of the
colliding nuclei -- may lead to a non-zero contribution to the directed flow
of charged particles and possibly could be disentangled by measuring the directed
flow of particles with equal mass but opposite
electric charge. Furthermore, one might have direct access to the positive
and negative initial charge density by measuring charge differential flow.
In this respect it is of great interest to compare the
$v_1$ rapidity/pseudorapidity dependencies and transverse momentum dependence
for identified hadrons that differ  by the electric charge, $v_1^+$ and $v_1^-$,
and to provide quantitative predictions for the differential observables.

Very recently, asymmetric Cu+Au collisions have been performed at
RHIC and the directed flow for identified particles has been measured
by the PHENIX Collaboration~\cite{Hu10,Io,Io13}.    In
Cu+Au collisions, a substantial electric field directed from a
colliding Au nucleus to the Cu nucleus is generated in the overlap
region (see below).  This happens only when the colliding two nuclei
carry a different number of electric charge. This electric field will
induce an electric current in the matter created after the collision,
resulting in a dipole deformation of the charge distribution. The time
evolution of the system is known to be dominated by the strong radial
flow, which is an outward collective motion of the medium.  Henceforth
the charge asymmetry formed in the early stage is frozen. Thus, it is
argued that the charge-dependent directed flow of the observed hadrons is
sensitive to the charge dipole formed at the early stage \cite{HHH12},
which reflects the electric conductivity of the QGP~ \cite{Ca11}
at early times. In symmetric collisions some other charge splitting
may appear, its magnitude appears to depend strongly on the
actual distance between the pion emission site and the spectator
system and bring new, independent information on the space-time
evolution of pion production~\cite{RS14}.

We briefly recall that the early magnetic field $eB_y$ \cite{GKR14} also may
lead to the induction of charged currents in the system. These currents
result in a charge-dependent directed flow  that is odd in
rapidity and odd under charge exchange and has to be added to the effects
studied in this work due to the electric field $eE_x$.

The new goal of this paper is the study of the
charge-dependent directed flow in terms of the PHSD model for a quantitative
specification/prediction of the $v_1$ splittings to be expected in two different
scenarios. We will consider the production of $\pi^+, \pi^-, K^+,
K^-$ mesons and $p, \bar{p}$ baryons in Cu+Au collisions at
$\sqrt{s_{NN}}=200$~GeV  taking into account the retarded
electromagnetic field that is created dominantly by the proton spectators in the
very early  collision \cite{VT11}. As primary observables we will address
the differential directed flow
$v_1(\eta, p_t)$ of these hadrons and explore the experimental perspectives.

\section{Reminder of the PHSD model}

The dynamics of partons, hadrons, and strings in relativistic
nucleus-nucleus collisions is treated within the
parton-hadron-string dynamics (PHSD) approach. The PHSD model is a
covariant dynamical approach for strongly interacting systems
formulated on the basis of Kadanoff-Baym
equations~\cite{JCG04,CB09} or off-shell transport equations in
phase-space representation, respectively. In the Kadanoff-Baym
theory the field quanta are described in terms of dressed
propagators with complex selfenergies. Whereas the real part of
the selfenergies can be related to mean-field potentials of
Lorentz scalar, vector or tensor type, the imaginary parts provide
information about the lifetime and/or reaction rates of time-like
particles~\cite{Ca09}. Once the proper complex self-energies of
the degrees of freedom are known, the time evolution of the system
is fully governed by off-shell transport equations for quarks and
hadrons (as described in Refs.\cite{JCG04,Ca09}). The PHSD model
includes the creation of massive quarks via hadronic string decay
- above the critical energy density
$\varepsilon_c \approx $0.5 GeV/$fm^3$ - and quark
\begin{figure*}[thb]
\includegraphics[height=5.50truecm] {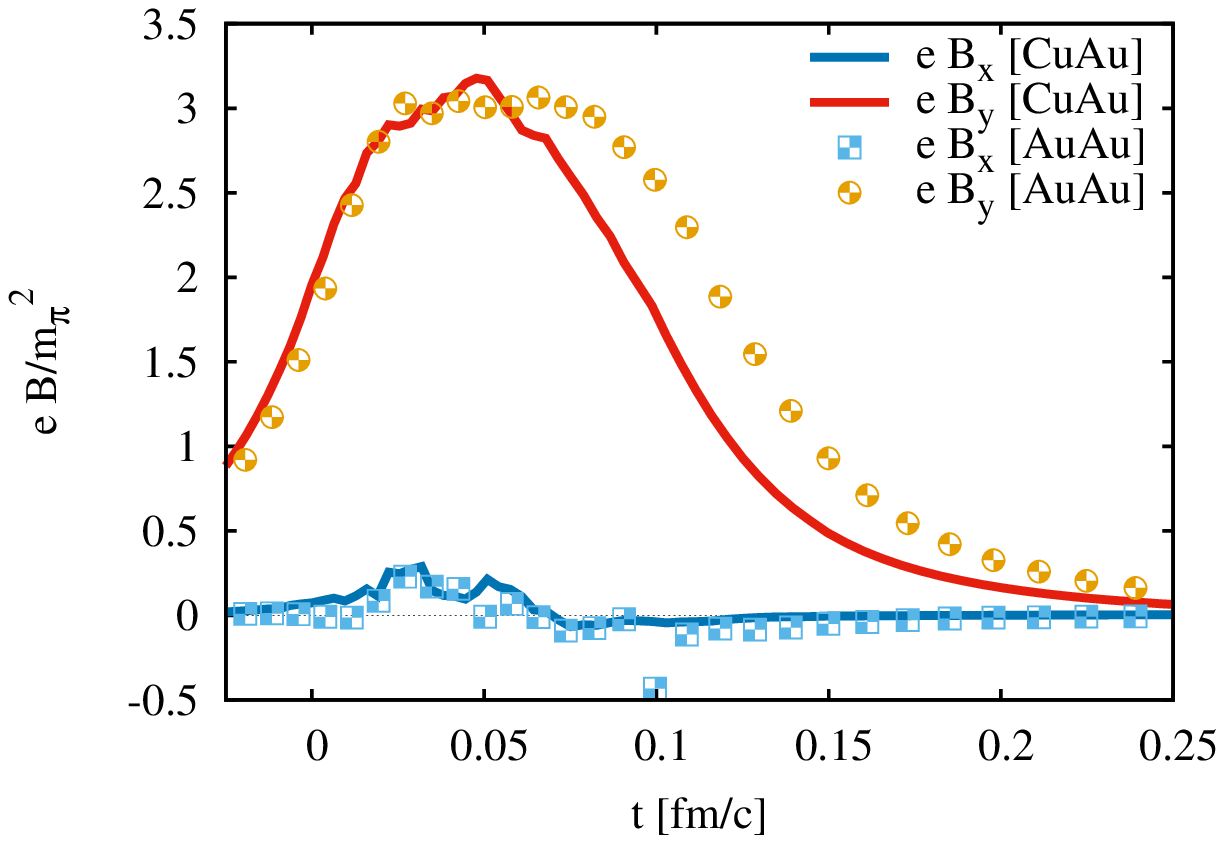}
\hspace{3mm}
\includegraphics[height=5.50truecm] {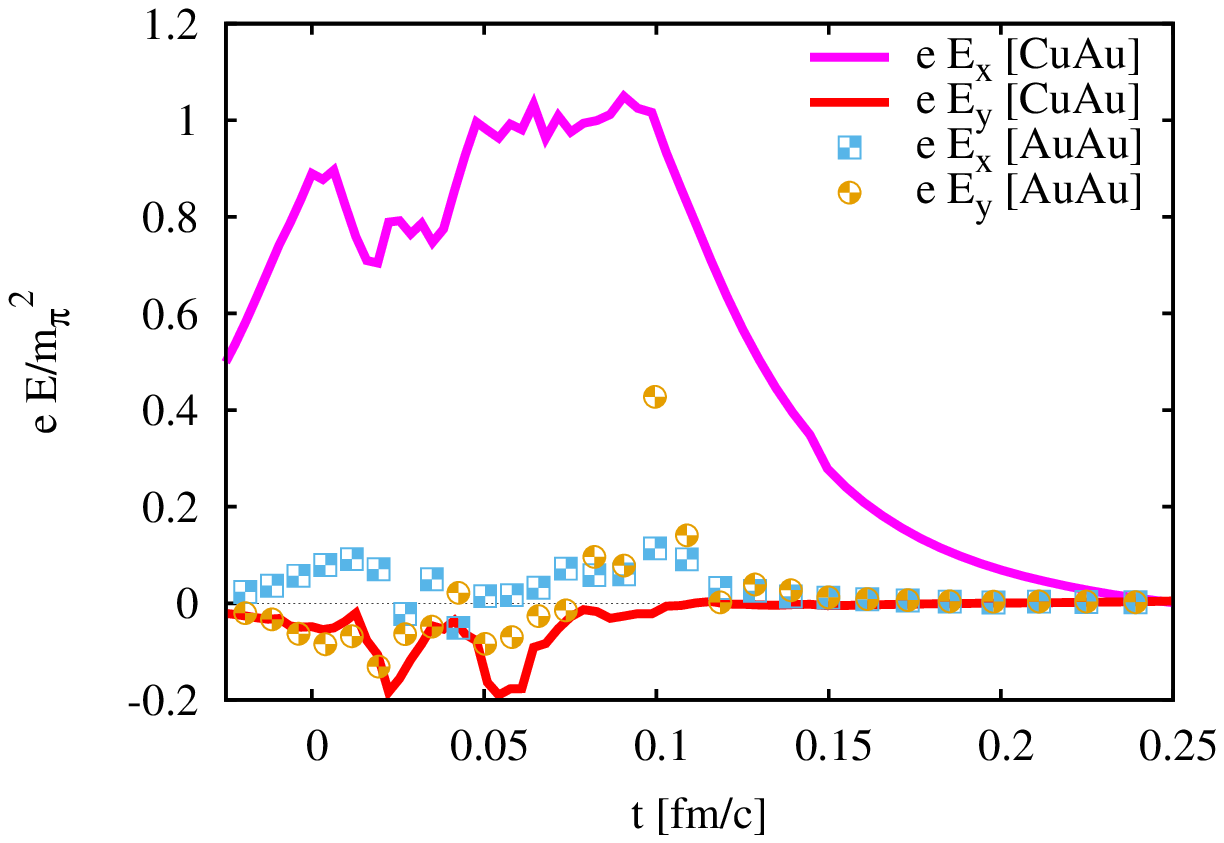}
\caption{Time evolution of event-averaged components of the magnetic
(l.h.s.) and electric (r.h.s.) fields in the center of the overlap
region of  colliding Cu+Au (solid lines) and Au+Au (dotted lines)
systems at $\sqrt{s_{NN}}=$200 GeV and $b=7$~fm. The distributions
are averaged over  70 events.} \label{E0B0-t}
\end{figure*}
\begin{figure*}[thb]
\includegraphics[height=6.50truecm]{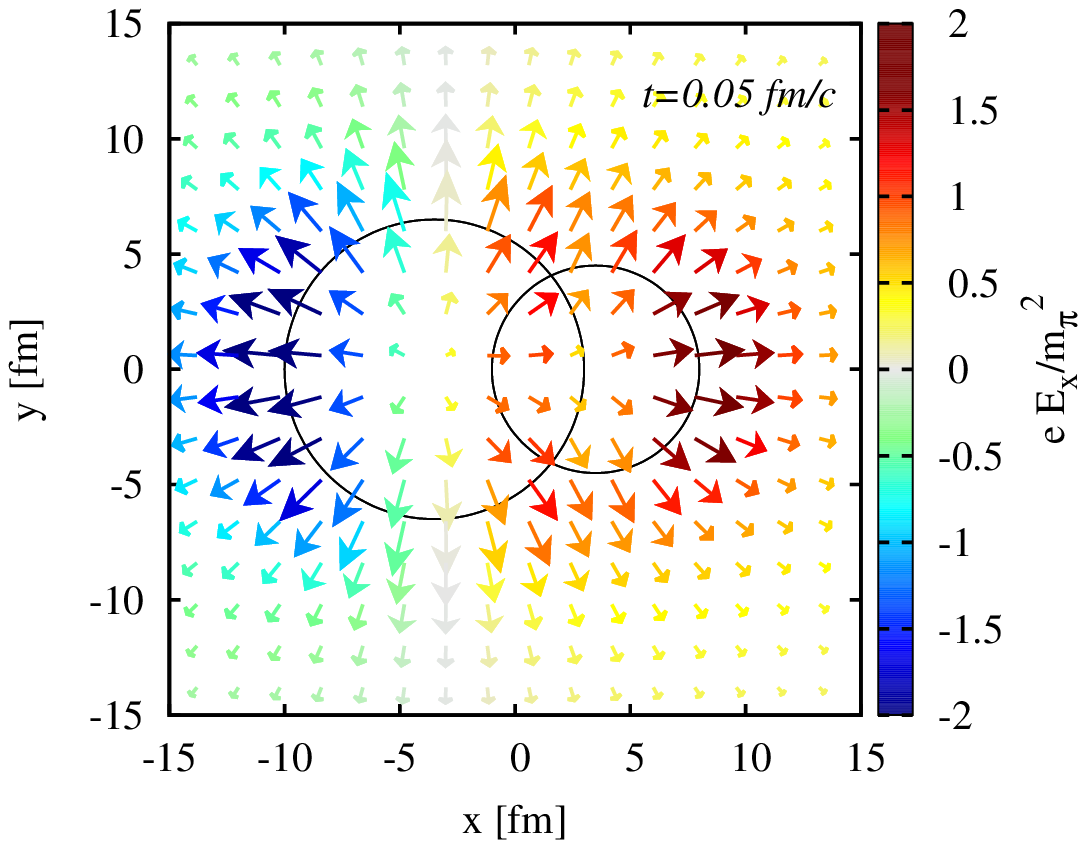}
\includegraphics[height=6.50truecm]{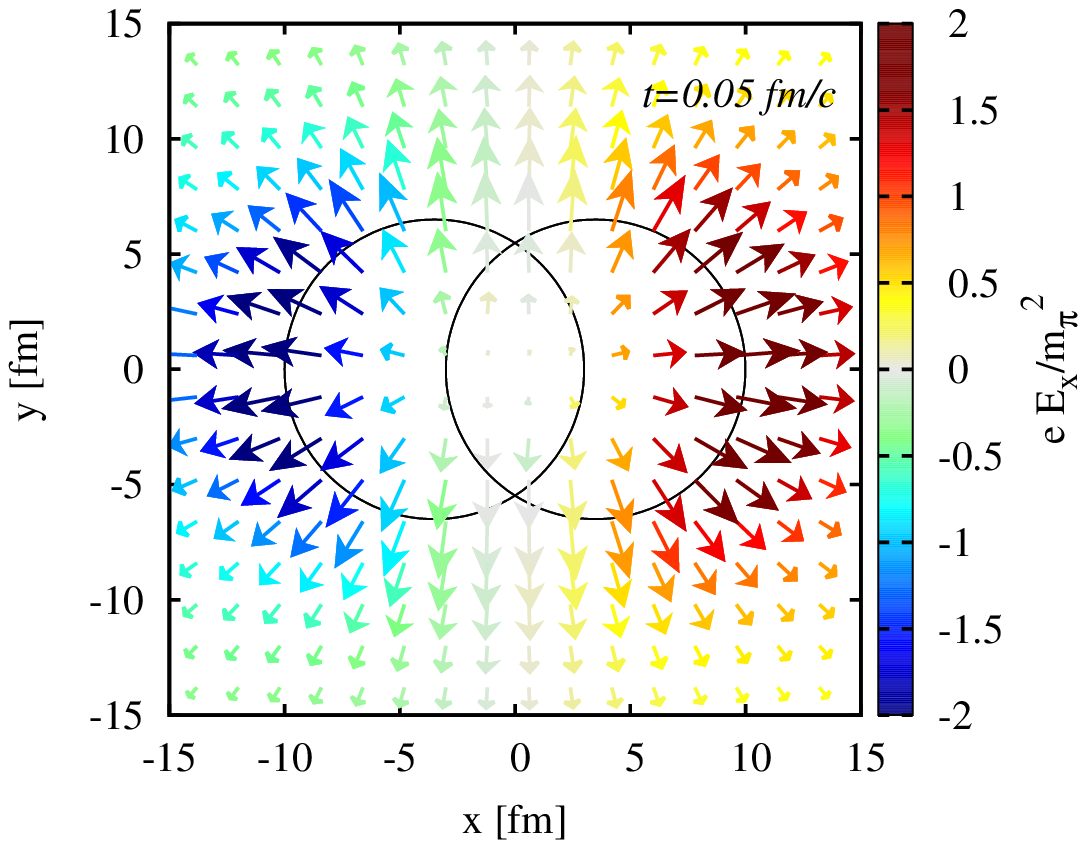}
\caption{Event-averaged electric field in the transverse plane for a
  Cu+Au (left panel) and Au+Au (right panel) collision at 200 GeV at
  time $t=$0.05~fm/c for the impact parameter $b=$7~fm. Each vector
  represents the direction and magnitude of the electric field at that
  point.  }
\label{field-diag}
\end{figure*}
fusion forming a hadron in the hadronization process. With some
caution, the latter process can be considered as a simulation of a
crossover transition since the underlying equation of state (EoS) in PHSD is a
crossover~\cite{Ca09}. At energy densities close to the critical
energy density $\varepsilon_c$, the PHSD describes a coexistence of the
quark-hadron mixture. This approach allows for a simple and
transparent interpretation of lattice QCD results for
thermodynamic quantities as well as correlators and leads to
effective strongly interacting partonic quasiparticles with broad
spectral functions. For a review on off-shell transport theory we
refer the reader to Ref.~\cite{Ca09}; PHSD model results and their
comparison with experimental observables for heavy-ion collisions
from the lower super-proton-synchrotron (SPS) to RHIC energies can
be found in Refs.\cite{Ca09,Ko12,Li11,To12}. In the hadronic
phase, i.e. for energies densities below the critical energy
density $\varepsilon_c$, the PHSD approach is identical to the
Hadron-String-Dynamics (HSD) model~\cite{EC96,CB99,CBJ00}.

In the PHSD the initial kinetic energy is converted in hard nucleon-nucleon
collisions to string like configurations via PYTHIA 6.4 and these
'strings' decay to partonic quasiparticles with spectral functions in
line with the dynamical quasiparticle model (DQPM, see \cite{Ca09}) if
 the energy density is above
critical ($\approx 0.5$ GeV/fm$^3$). The initial conversion of energy
happens during roughly 0.15 fm/c at the top RHIC energy when the nuclei
pass though each other. At this time the energy density in between the
leading baryons is very high due to the fact that the spacial
volume is very small and  $\sim$ 0.3 fm in longitudinal extension; the transverse
contribution to this volume is given by the overlap area.
Due to the Heisenberg uncertainty relation this
energy density cannot be specified as being due to 'particles' since
the latter may form only much later on a timescale of their inverse
transverse mass (in their rest frame). More specifically, only  jets
at midrapidity with transverse momenta above $p_T =$ 2 GeV are expected to
appear at $t \approx 0.1$ fm/c while a soft parton with transverse
momentum $p_T = 0.5$ GeV should be formed after $t \ge$ 0.5 fm/c.
Although it is not clear what the actual nature of the degrees of freedom is
in this initial state, there will naturally be a small amount of electric
charges due to charge conservation.  On the other hand, if a large amount
of electric charges (from the conversion of energy to quarks and antiquarks)
are present in the very beginning
of the reaction, then there should be observable signals from this early
electric accelerator. It is our aim to quantify within PHSD these possible
signals and to provide robust predictions.

We use here the PHSD version where the creation of electromagnetic fields
and particle transport in these fields are taken into account by means of the
 retarded Li\'enard-Wiechert potentials~\cite{VT11}. Only the source of the
spectator protons is considered since this source is  dominant at the initial
stage when target and projectile spectators are close to each other.
By the time of about 1 fm/c - after contact of the nuclei - the electromagnetic
fields drop down by three orders of magnitude and become comparable with the
field from the participants \cite{VT11}. This offers the very specific
property of the early electric field to check experimentally if electric
charges are already present at this instant.

The time evolution of  transverse electromagnetic
field components is compared between asymmetric Cu+Au (solid lines)
and symmetric Au+Au systems (dotted lines) in Fig.~\ref{E0B0-t}
where the l.h.s. displays the magnetic field components
and the r.h.s. the electric ones. The maximal values of the magnetic
field components $\langle eB_y\rangle$ are on
the level of a few $m_\pi^2$ being comparable for both
systems. For the symmetric case the results are in
agreement with our earlier results in~\cite{VT11}. The
electric field components also agree with the earlier results
for symmetric collisions~\cite{VT11} but in case of the Cu+Au
reaction the $\langle eE_x\rangle$ component is by a
factor of $\sim$5 larger than that for symmetric Au+Au collisions
at the same energy~\cite{VT11}. This strong electric field $eE_x$
is only present for about 0.25 fm/c during the overlap phase of
the heavy ions and will act as an electric accelerator
on charges that are present during this time. Note that when charges
appear only later together with the formation of soft partons
($t \ge $ 0.5 fm/c) there will be no corresponding charge separation
effect on the directed flow! In the case of symmetric collisions
it was noted that $\langle E_x\rangle \approx \langle
B_y\rangle$~\cite{VT11,BS12}. This approximate equality is broken for
asymmetric Cu+Au collisions where $\langle eB_y\rangle > \langle
eE_x\rangle$.

Fig.~\ref{field-diag}, furthermore, shows the distribution in the
strength and direction of electric field components
for off-central Cu+Au and Au+Au collisions. This snapshot is made for
the time when both nuclear centers are in the same transverse
plane. This condition corresponds to different times for the two
systems considered  which is confirmed by a shift of the component
$\langle eB_y\rangle$ in time (cf. l.h.s. of Fig. 1) where
the maximum is reached earlier in Cu+Au collisions. Here we take
$t\sim$0.05 fm/c in view of Fig. 1. Note that in Cu+Au collisions a
significant electric field $eE_x$ is generated in the overlap region of the two
nuclei in $x$-direction, i.e. directed towards the lighter copper nucleus.
The situation is different  in collisions of nuclei of the same
size~\cite{VT11,BS12,DH12} as illustrated in Fig.~\ref{field-diag}
(r.h.s.).  In symmetric collisions like Au+Au or Cu+Cu, the
event-averaged electric field does not show a
preferential direction and the magnitude of the electric fields
generated in each event is lower, too.

This strong electric field $e E_x$ towards the Cu nucleus at the
early stage induces an electric current in the medium (if electric
charges are present). As a result, the charge distribution is
modified and a charge dipole is formed~\cite{HHH12}. In central
Cu+Au collisions the Cu nucleus is completely embedded within the
Au-nucleus and due to the absence of Cu spectators no sizeable
electric current is formed. We note in passing that the
electric field sharply drops after $t\gsim $0.25 fm/c in free space,
while in  conducting matter the time dependence of the field
strength is flattening out and reaches some plateau even up to
$t\sim$ 10 fm/c~\cite{Tu13}. The level of this plateau is
proportional to the electric conductivity $\sigma$ and therefore the
conductivity effect could be sizeable in the case of a weakly
interacting QGP. However, the electric conductivity - as evaluated
within PHSD in a finite box with periodic boundary conditions - is
much lower and comparable to lattice QCD results for temperatures
from 170 MeV to 250 MeV. For more details and explicit comparisons
we refer the reader to Ref.~\cite{Ca11}.  

Now the question is: what is the maximal strength of this induced
current and how to see that experimentally?

\section{PHSD predictions for the directed flow in Cu+Au collisions}

It is widely recognized that fluctuations in the initial geometry of
colliding nuclei are very important. At fixed impact parameter,
depending on the location of the participant nucleons in the nucleus
at the time of the collision, the actual shape of the overlap area may
vary: the orientation and eccentricity of the ellipse defined by the
participants fluctuates from event to event, i.e. the reaction plain
fluctuates, too. We recall that  taking into account these
fluctuations in the initial geometry, the PHSD model reasonably
describes the rapidity dependence $v_1(y)$ of the
directed flow of charged hadrons including their slopes at  midrapidity
as well as the flow dependence on beam energy for symmetric Au+Au
collisions~\cite{KBC12}.

Electromagnetic effects in particle emission clearly imply the
dependence of specific observables on a particle charge, in particular
also for particles of the same mass (like e.g. $\pi^+$ and $\pi^-$
mesons). For specific components of the electromagnetic field
contributing to pion directed flow (different from that in asymmetric
collisions), such a charge dependence - charge splitting - was
predicted even for symmetric nuclear combinations in Refs.
\cite{HHH12,GKR14}. In a simplified treatment of \cite{HHH12} the
effect is reduced to the relative distance between spectator blobs,
which is a parameter to be
\begin{figure*}[bht]
\includegraphics[height=5.50truecm] {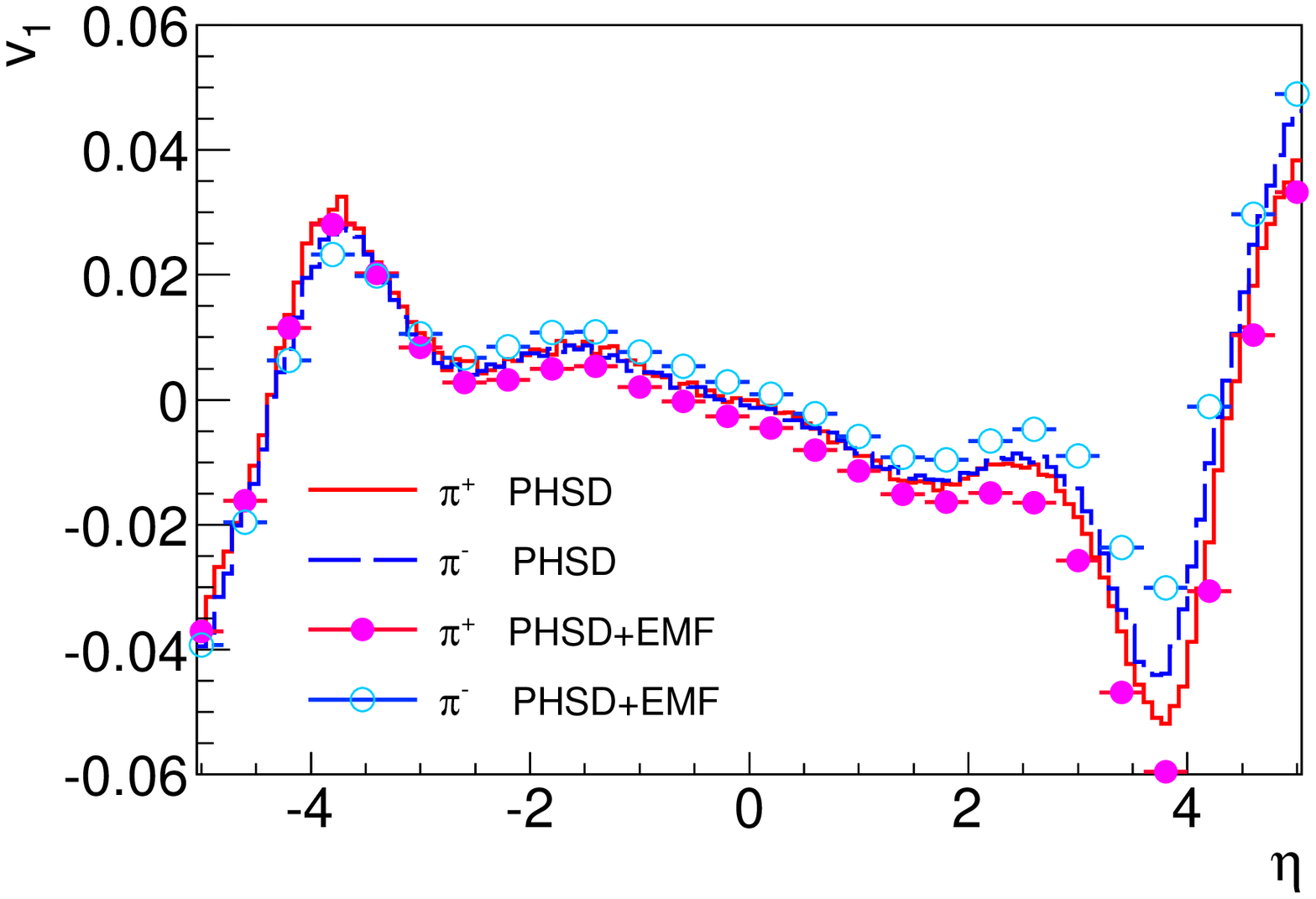}
\hspace{3mm}
\includegraphics[height=5.50truecm] {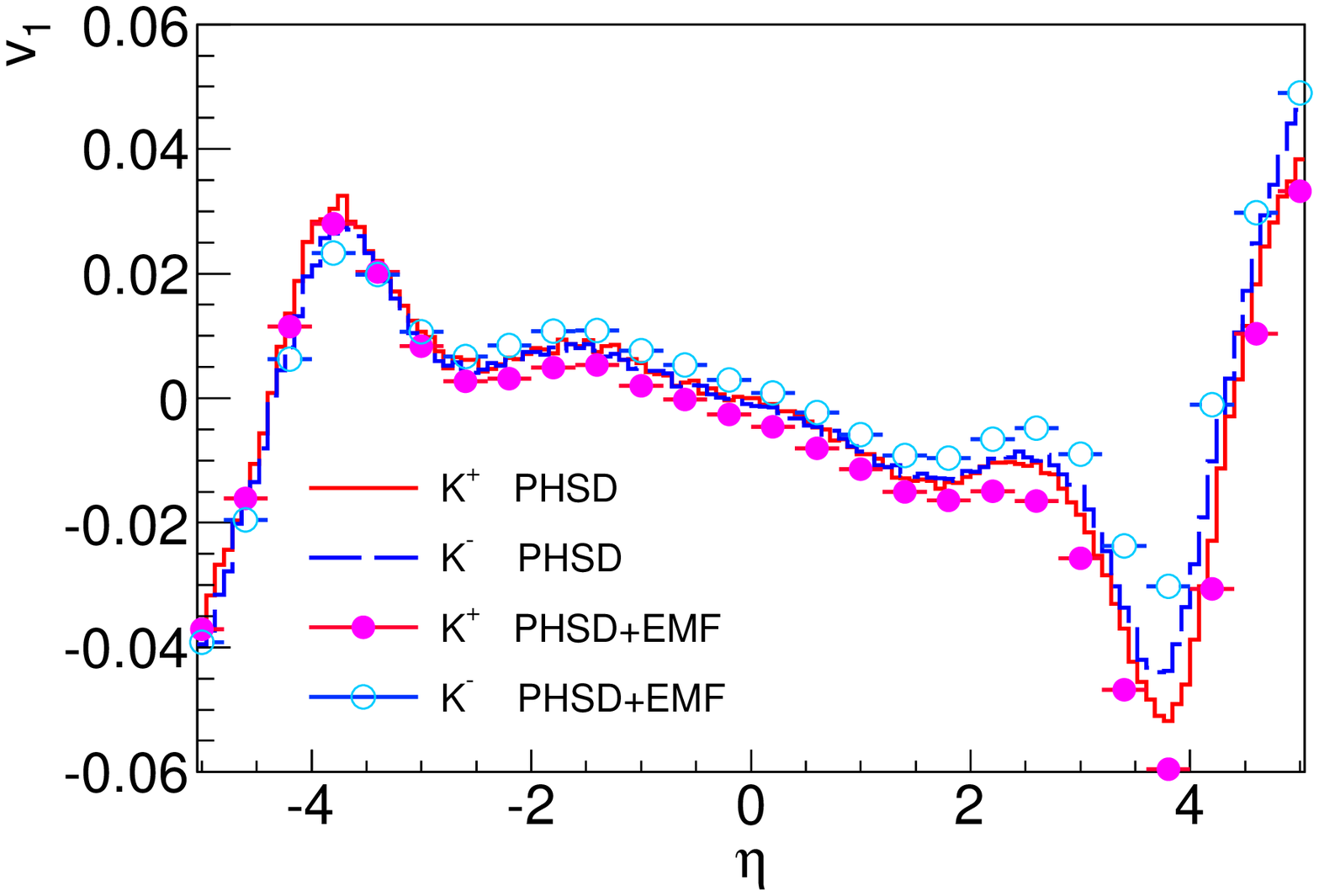}
\caption{Rapidity dependence of the directed
  flow of pions and kaons created in off-central Cu+Au collisions at
  $\sqrt{s_{NN}}=200$ GeV.  The PHSD results for positive pions and
  kaons $v_1^+$ and for negative ones $v_1^-$ are plotted by the solid and
  dashed histograms, respectively. Results  by including the effect of the
  electromagnetic field on early charges are marked  by filled and empty
  circles for $v_1^+$ and $v_1^-$, respectively.  }
  \label{v1-pi}
\end{figure*}
fixed by data in order to achieve an agreement with a
particular experiment and therefore it is not robust. The
importance of precise experimental data on directed flows
measured separately for positive and negative pion charges
$v_1^{\pi+},v_1^{\pi-}$ becomes clearly evident.

The results of the STAR Collaboration on the directed
flow of protons, antiprotons and pions in the Au+Au collision
energy range from $\sqrt{s_{NN}}=$7.7 up to 200 GeV have been
recently published~\cite{v1STAR}. These data include measurements
of $v_1$ for $p,\bar{p}, \pi^+$ and $\pi^-$   for
seven collision energies in the c.m.s. rapidity range of $|y|<$1.
The comparison of positive and negative pion directed flow at the
$\sqrt{s_{NN}}=$7.7 GeV (but not at higher energies) in
intermediate centrality (10-40\%) Au+Au collisions displays a
clear splitting of $v_1^{\pi+}$ and $v_1^{\pi-}$ , with
$v_1^{\pi+}<v_1^{\pi-}$ at positive rapidity. As noted
in~\cite{GKR14}, this appears  consistent with the expectation
of a specific charge-dependent component of the directed flow
induced by electromagnetic effects but this connection should be
checked  independently.

Similarly to the symmetric case, the directed flow of hadron
distributions in asymmetric Cu+Au collisions is defined as
\begin{equation}
v_1=\langle\cos (\phi- \Psi_{RP})\rangle
\end{equation}
where azimuthal angles $\phi$ are measured with respect to the
transverse direction of the reaction plane $\Psi_{RP}$. 
Ideally, the reaction plane $\Psi_{RP}$ is defined by the vector of
the impact parameter and the beam direction. It is found that the
magnitude of $v_1$ is correlated  with the determination of the
reaction plane. Here we associate $\Psi_{RP}$ with the participants
of the Au-nucleus. For a comparison of $v_1$ distributions resulting
from different definitions of the reaction plane we refer the reader
to Ref.~\cite{WMZS14}.  The results presented in Fig.~\ref{v1-pi}
are calculated for identified $\pi^\pm$ and $K^\pm$ mesons taking
into account the  spectator electromagnetic field as well as its
influence on quasiparticle transport in both partonic and hadronic
phases.  The filled and empty circles result from calculations that
assume all partonic charges to be present immediately after string
dissolution to 'partons'. The dashed and solid histograms are
obtained in the other scenario when  the early electric acceleration
is discarded or when the charges are delayed to their formation time
and therefore appear together with the formation of the charged
partons (quarks and antiquarks).  These distributions as well as
those presented below are calculated for the impact parameter range
4.7$\le b \le$9.5~fm which corresponds to (10-40)\% centrality. The
number of simulated events amounts to $2\times 10^6$ in the case of
including the electromagnetic field and without it. The
pseudorapidity dependence of $v_1$ exhibits a negative slope around
mid-$\eta$. In addition, $v_1(\eta)$ displays a dipole-like
asymmetry between the forward and backward rapidity distributions.
The magnitude of $v_1$ at the forward pseudorapidity (the Au-like
side) is higher than that at the backward one  (the Cu-like side).
Due to this nuclear asymmetry the pseudorapidity dependence of
$v_1(\eta)$ for every hadron does not go through zero at midrapidity
($\eta=0$), as follows from
 Table~\ref{tab1}. All the slopes at this point $dv_1/d\eta(\eta=0)$
 are negative and small as presented in Table~\ref{tab2}.
An even stronger charge asymmetry of $v_1$ in forward and backward
rapidities is seen at larger pseudorapidity ($|\eta|\ge$ 2.5),
with the magnitude of $v_1$ at large backward pseudorapidity
 (Cu-like rapidity)  being lower than that at large forward
pseudorapidity (Au-like rapidity).

\begin{table}
\begin{center}\begin{tabular}{|c|c|c|}
\hline
hadron \ & \ $v_1(\eta=0)$ PHSD & \ $v_1(\eta=0)$ PHSD+EMF \    \\
\hline
$\pi^+$  & (-0.68 $\pm$ 0.140)e-03 \ &  (-3.87 $\pm$ 0.133)e-03  \\
$\pi^-$  & (-1.09 $\pm$ 0.136)e-03 \ &  ( 1.71 $\pm$ 0.129)e-03  \\
$K^+$    & (-1.37 $\pm$ 0.312)e-03 \ &  (-6.33 $\pm$ 0.311)e-03  \\
$K^-$    & (-2.11 $\pm$ 0.327)e-03 \ &  ( 2.43 $\pm$ 0.327)e-03   \\
$p$      & (-2.82 $\pm$ 0.443)e-03 \ &  (-8.37 $\pm$ 0.419)e-03 \\
$\bar{p}$& (-1.58 $\pm$ 0.504)e-03 \ &  ( 4.94 $\pm$ 0.479)e-03 \\
\hline
\end{tabular} \end{center}
\caption{ The directed flow at mid-$\eta$, $v_1(\eta=0)$, for
Cu+Au ($\sqrt{s_{NN}}$= 200 GeV) collisions at $4.7<b<9.5$ fm. }
\label{tab1}
\end{table}
\begin{table}
\begin{center}\begin{tabular}{|c|c|c|}
\hline
hadron \ & \ $dv_1/d\eta$ PHSD & \ $dv_1/d\eta$ PHSD+EMF \    \\
\hline
$\pi^+$  & (-5.26 $\pm$ 0.340)e-03 \ &  (-6.39 $\pm$ 0.299)e-03  \\
$\pi^-$  & (-5.68 $\pm$ 0.330)e-03 \ &  (-6.18 $\pm$ 0.291)e-03  \\
$K^+$    & (-5.28 $\pm$ 0.751)e-03 \ &  (-4.97 $\pm$ 0.750)e-03  \\
$K^-$    & (-6.24 $\pm$ 0.789)e-03 \ &  (-7.26 $\pm$ 0.789)e-03  \\
$p$      & (-14.9 $\pm$ 1.07 )e-03 \ &  (-6.39 $\pm$ 0.299)e-03 \\
$\bar{p}$& (-3.83 $\pm$ 1.22 )e-03 \ &  (-6.18 $\pm$ 0.291)e-03 \\
\hline
\end{tabular} \end{center}
\caption{ Estimated slope parameters $dv_1/d\eta(\eta=0)$ for
Cu+Au ($\sqrt{s_{NN}}$=200 GeV) collisions at $4.7<b<9.5$ fm. Parameters of
the $\eta$ distributions are defined by a linear fit for $|\eta|<0.7$.
} \label{tab2}
\end{table}

\begin{figure}[thb]
\includegraphics[height=5.50truecm] {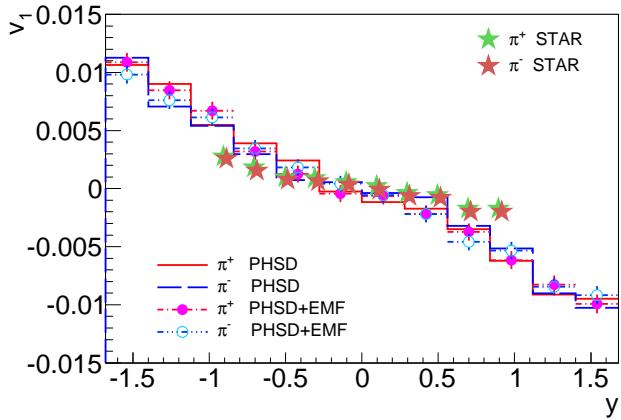}
\caption{Rapidity dependence of the pion directed flow  in Au+Au
collisions at $\sqrt{s_{NN}}=$200 GeV. The solid histogram is calculated
in PHSD for $\pi^+$ and the dashed one  for  $\pi^-$ mesons. Results  from PHSD
  by including the effect of the
  electromagnetic field on early charges are marked  by filled and empty
  circles for $v_1^+$ and $v_1^-$, respectively. Experimental
  data points (stars) are from Ref.~\cite{v1STAR}.  }
  \label{v1-pi-AuAu}
  \end{figure}

 In Fig.~\ref{v1-pi-AuAu} the directed $\pi^+$ $v_1^{\pi+}(y)$ and $\pi^-$
 $v_1^{\pi-}(y)$ flows from PHSD are contrasted to
  the symmetric Au+Au case. As is seen there
 is practically no dependence of the $v_1(y)$ distributions on the pion
 charge as well as on the electromagnetic field. Note that similar PHSD
 results have been obtained earlier for the same colliding system from the
 analysis of elliptic flow~\cite{Ko12}. In agreement with this
 finding, the $\pi^+$ and $\pi^-$ meson spectra  measured in Au+Au
 collisions coincide with each other~\cite{v1STAR}.  Accordingly, the
 charge-dependent asymmetry of the directed flow in Cu+Au collisions
 originates from the electric field asymmetry induced by the asymmetry in
 charge of the initial colliding nuclei but not as a particularity of the
 observable under consideration.

\begin{figure}[thb]
\vspace*{1mm}
\includegraphics[height=5.50truecm] {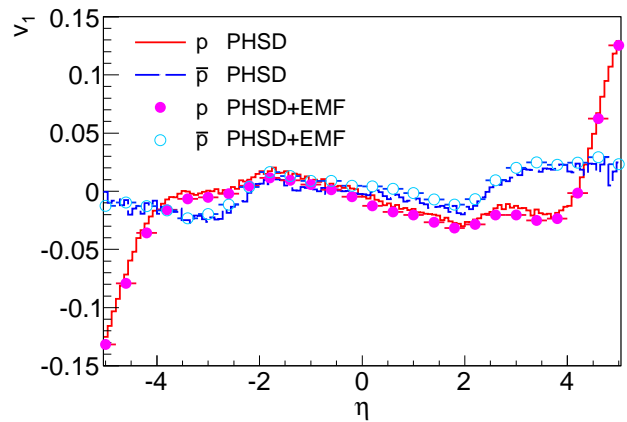}
\caption{Rapidity dependence of the directed
  flow of protons and antiprotons in Cu+Au collisions at
  $\sqrt{s_{NN}}=$200 GeV.  The notation is the same as
  in Fig.~\ref{v1-pi}.  }
  \label{v1-p}
  \end{figure}

The rapidity dependence of the directed flow $v_1(\eta)$ for $p/{\bar p}$
is presented in Fig.~\ref{v1-p}. For antiprotons, the $v_1(\eta)$ is
comparatively flat with a weak increase towards large pseudorapidities.
The proton and antiproton results are close to each other at
$|\eta| \lsim$2.5 but strongly differ at higher rapidities due to
contributions from the fragmentation process. The shape of the proton
directed flow distribution $v_1(\eta)$ resembles that for the created
mesons (cf. Fig.~\ref{v1-pi}) since the mesons and
baryons/antibaryons emerge dominantly by hadronization from the same
partonic medium.
\begin{figure*}[thb]
\includegraphics[height=5.50truecm] {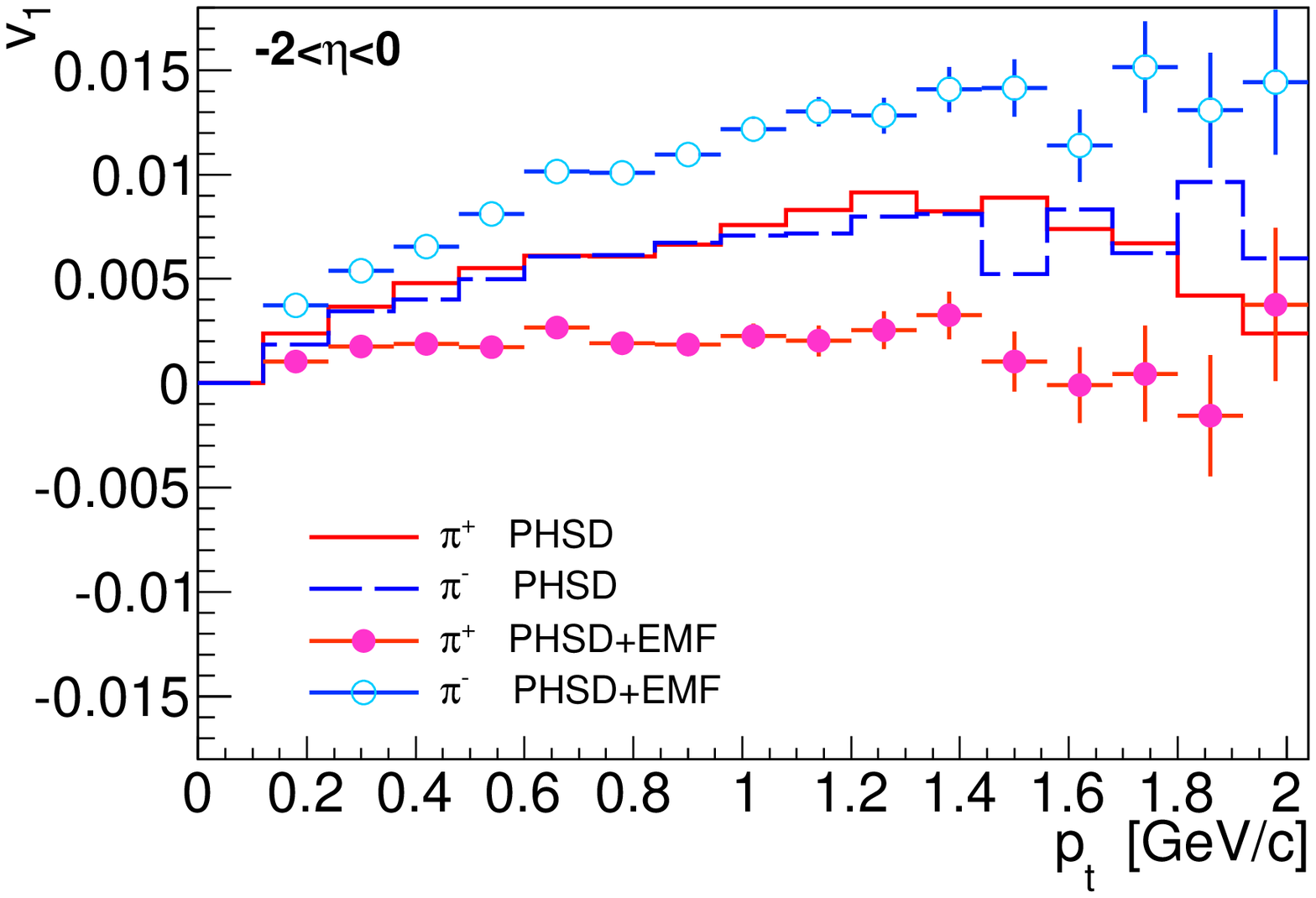}
\hspace{3mm}
\includegraphics[height=5.50truecm] {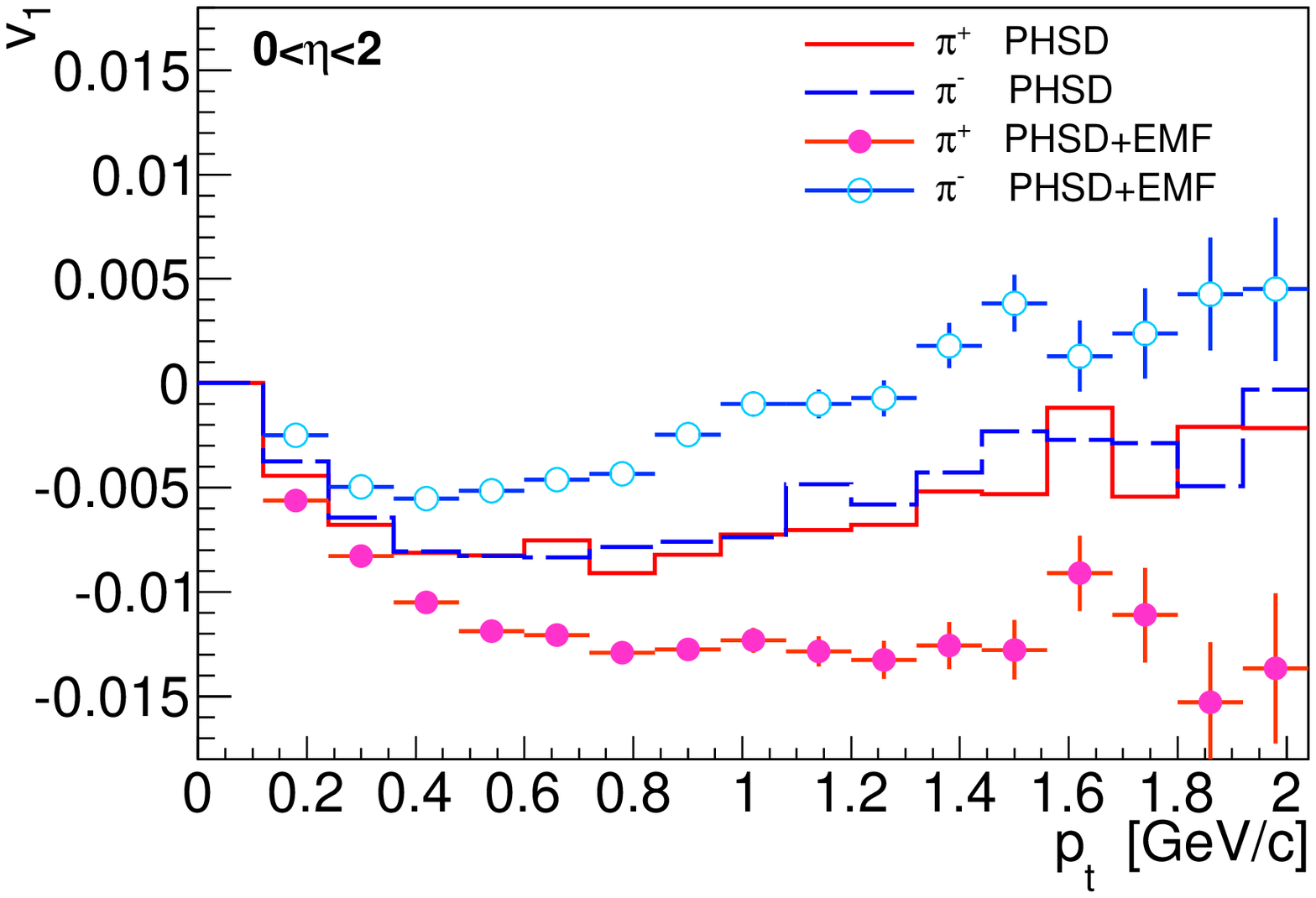}
\includegraphics[height=5.50truecm] {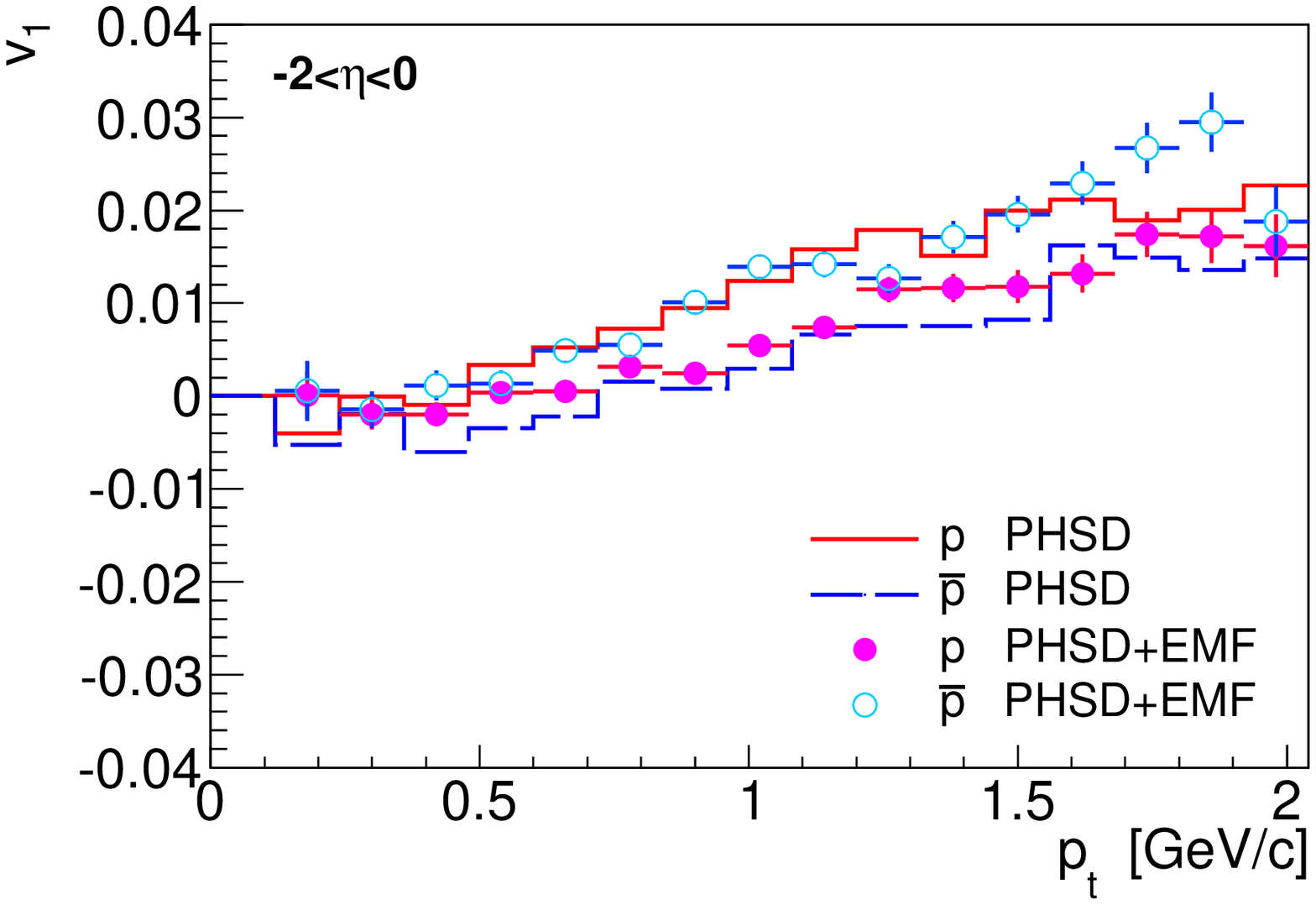}
\hspace{3mm}
\includegraphics[height=5.50truecm]{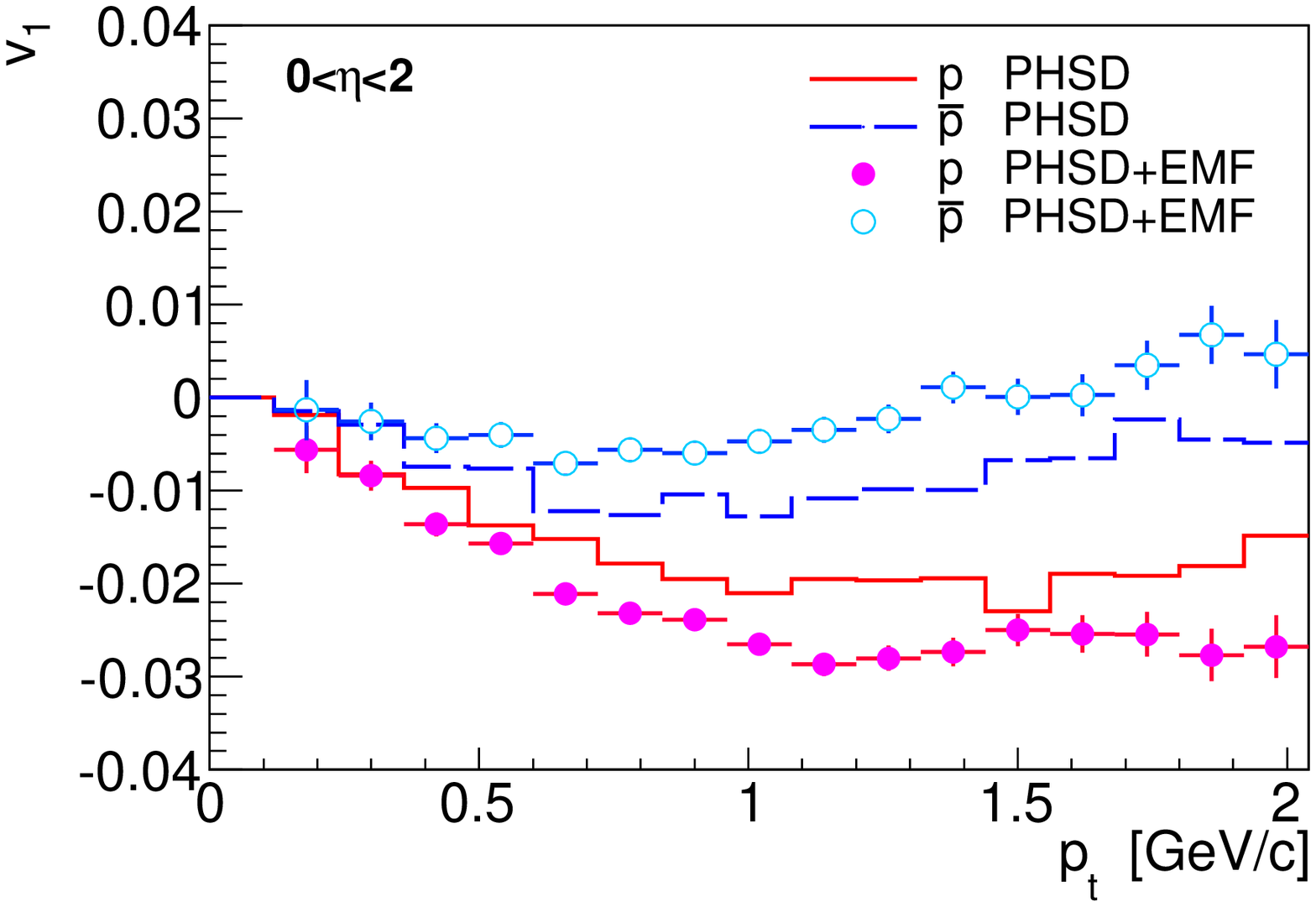}
 \caption{Transverse momentum dependence of the directed flow for $\pi^\pm$
 (upper panels) and protons and antiprotons (bottom panels) produced in
 peripheral Cu+Au collisions for forward (right panels) and backward (left
   panels) emitted hadrons at $\sqrt{s_{NN}}=$200 GeV.
   The notation is the same as in Fig.~\ref{v1-pi}.  }
\label{v1-pt}
\end{figure*}
One should note that the observed differences between the directed
flow of protons and antiprotons could give information also on the different
rate of baryon stopping on the Cu and Au side of the fireball. Moreover, the
reaction plane correlations in the fireball are modified in asymmetric
collisions with stronger correlations between odd and even order
event planes \cite{BBM11,BLO11}.

Fig.~\ref{v1-pt} shows the PHSD results for transverse momentum
distributions of the directed flow. It is seen that the $v_1(p_T)$
functions for $\pi^+$ and $\pi^-$ mesons coincide with each other if
electromagnetic forces are neglected (histograms in the upper panel of
Fig.~\ref{v1-pt}) or if partonic charges appear only after
$t \ge$ 0.5 fm/c. In this case the $v_1^\pi$ is negative in the forward
direction and positive at negative angles $-2<\eta<0$. The inclusion
of the early electromagnetic field noticeably changes these $p_T$
distributions increasing $v_1$ for negative pions and decreasing $v_1$
for positive pions. In both cases the splitting effect is clearly seen
but the shape of $v_1^\pi(p_T)$  also depends on the pion electric
charge and pseudorapidity interval. For the backward $\pi^-$ emission
the directed flow $v_1$ grows monotonically with momentum while
$v_1^{\pi+}(p_T)\approx 0$ for $p_T\lsim $1.4 GeV and then decreases
with $p_t$. As to the forward angles $0\le \eta\le$2, the directed flow $v_1^-$ increases
also but exhibits a wide negative minimum in the region of
$p_T\approx$0.3-0.5 GeV/c. The directed flow of positive pions is seen to
decrease and then to flatten at $p_T\ge$1 GeV/c, being negative in the
whole transverse momentum range.

The directed flow $v_1(p_T)$ for protons and antiprotons shown in
Fig.~\ref{v1-pt} (bottom panels) displays a strong asymmetry in the forward and backward
pseudorapidities, having a larger $v_1(p_T)$ splitting at forward  (the
Cu-side) pseudorapidity than that at backward (the Au-side)
pseudorapidity. Note that even without taking into account the
electromagnetic effect there is a difference in $v_1$ between
$p$ and $\bar{p}$ due to different interaction mechanisms of these
hadrons (see bottom panels in Fig.~\ref{v1-pt}). In the baryonic case
the influence of the created electromagnetic field is not so large as
for pions and is more easily seen in the forward direction
$0\le\eta\le$2. The momentum dependence of
$v_1(p_T)$ in the backward direction is a positive monotonously
increasing function but in the forward direction $v_1(p_T)\lsim$0 with
a weak wide minimum at $p_T\sim$0.7 GeV/c for the case of
antiprotons. The distribution $v_1(p_T)$ for protons is similar to that
for $\pi^+$: it decreases till $p_T\sim$1~GeV/c and then flattens.

We finally note in passing that when including also the electromagnetic
fields as induced by participant charges in addition to those discussed
 before, the control PHSD calculations ($\sim 1.5\times 10^5$ events) on the
charge-dependent directed flow presented in this work only change within
the statistical accuracy.

\section{Conclusions}

Asymmetric Cu+Au collisions have been studied at the ultrarelativistic
energy $\sqrt{s_{NN}}=$200 GeV within the PHSD approach which is in
a reasonable agreement with available experimental data for symmetric
Au+Au collisions in the energy range from the SPS to the top RHIC
energy. The retarded electromagnetic field induced by
spectator protons has been calculated (and quantified) and its influence
on the quasiparticle transport has been  taken into account within two
model scenarios: i) if the initial field energy at times $\sim $ 0.1 fm/c after
contact is dominantly carried by (still unformed) charged quarks and antiquarks
and ii) if the quark/antiquark creation (and accordingly the presence of
electric charges) is delayed to their formation time.

Independent on the two scenarios we have explicitly demonstrated that in
peripheral collisions of Cu+Au an additional short-time electric field
$eE_x$ in the direction of the lighter nucleus emerges
which at top RHIC energies is essentially effective for the first 0.25
fm/c and is due to the asymmetry in the number of protons in the target
and projectile nuclei. This additional electric field is not present
in symmetric Au+Au collisions but can be used in asymmetric collisions
to study the electric response of the medium during
the passage time of the nuclei, i.e. also out-off equilibrium.
In fact, the detailed PHSD predictions for the charge differential
directed flow $v_1(\eta,p_t)$ in Cu+Au reactions
show that the scenario i) leads to clearly observable
effects in the directed flow of hadrons with the
same masses but different electric charges, for example $\pi^+ -
\pi^-$, $K^+ - K^-$, $p - {\bar{p}}$. These differences can be
studied by means of the charge-dependent pseudorapidity
dependence of $v_1(\eta)$. High precision data on the transverse
momentum dependence $v_1(p_t)$, furthermore, should clarify the situation and
differentiate between the two scenarios because in case of ii) there
are almost no early electric charges of low $p_t$ and accordingly no
effects from the early  electric acceleration.

The experimental observation of the splitting effect in
charge-dependent observables of the direct flow would be a direct
confirmation of the importance of electromagnetic fields
in relativistic heavy-ion collision dynamics and might shed further
light on the effective degrees of freedom in the very early
non-equilibrium phase and possibly also on the photon '$v_2$-puzzle' \cite{Elena14}.

\section*{Acknowledgements}
We are grateful to E. L. Bratkovskaya and V. P. Konchakovski for useful
discussions. V. T. and V.V. acknowledge the Heisenberg-Landau grant support
and support within the HIC for FAIR framework.
The work of S.V.  is supported by the U.S. Department of
Energy Office of Science, Office of Nuclear Physics under Award
Number DE-FG02-92ER-40713.


\begin{thebibliography}{99}

\bibitem{VPS10} S. A. Voloshin, A. M. Poskanzer and R. Snellings, in
  Landolt-Boernstein New Series, I/23, p. 5-54, edited by R. Stock,
  Springer-Verlag, 2010.

\bibitem{So10} P. Sorensen, In Quark-Gluon Plasma 4, ed. by R. Hwa and
  X.N. Wang, World Scientific (2010).

\bibitem{PC05} A. Peshier and W.Cassing,
Phys. Rev. Lett. {\bf 94},   172301 (2005); M. Gyulassy and L. McLerran,
Nucl. Phys. {\bf A750}, 30 (2005).

\bibitem{RR97} W. Reisdorf and H.G. Ritter, Annu. Rev. Nucl. Part. Sci. {\bf 47}, 663 (1997).

\bibitem{HWW99} N. Herrmann, J. P. Wessels, and T. Wienold, Ann. Rev. Nucl. Part. Sci.
{\bf 49}, 581 (1999).

\bibitem{v1STAR} STAR Collaboration: L. Adamczyk, et al.,
  Phys. Rev. Lett. {\bf 112}, 162301 (2014).

\bibitem{v1KCT} V. P. Konchakovski, W. Cassing, Yu. B. Ivanov, and V. D. Toneev,
  Phys. Rev. {\bf C90}, 014903 (2014).

\bibitem{Yuri} Yu. B. Ivanov, V. N. Russkikh, and V. D. Toneev,
Phys. Rev. {\bf C73}, 044904 (2006).

\bibitem{Rischke} D. H. Rischke,
Nuclear Physics {\bf A610},  88 (1996).

\bibitem{HS05} H. St\"ocker,
Nucl. Phys. A {\bf 750}, 121 (2005).

\bibitem{CK06} L.-W. Chen and C. M. Ko, Phys. Rev. {\bf C73}, 014906 (2006).

\bibitem{Bo12} P. Bozek,
Phys. Lett. {\bf     B717}, 287 (2012).

\bibitem{STV14} B. Schenke, P. Tribedy, and R. Venugopalan,
arXiv:1403.2232.

\bibitem{Io13} A. Iordanova, for the PHENIX Collaboration,
  Journal of Physics: Conference Series {\bf 458},  012004( 2013);
  H. Nakagomi,
Riken Rad. Lab. Meeting (2014).

\bibitem{TY11}D. Teaney and L. Yan,
Phys. Rev. C {\bf 83} , 064904   (2011).

\bibitem{LO11} M. Luzum and J.-Y. Ollitrault,
Phys. Rev. Lett. {\bf 106 },
  102301 (2011)

\bibitem{HNM11} Md. Rihan Haque, M. Nasim, and B. Mohanty,
Phys. Rev. {\bf  C84}, 067901 (2011), arXiv:1111.5095 [nucl-ex].


\bibitem{BW10} P. Bozek and I. Wyskiel,
Phys. Rev. {\bf C81}, 054902   (2010).

\bibitem{ARG10} R. Andrade, A. Reis, F. Grassi, Y. Hama, T. Kodama and
  J.Y.  Ollitrault and W.L. Qian, Indian J. Phys., {\bf 84}, 1657
  (2010).

\bibitem{Hu10} S. Huang,
  arXiv:1210.5570.

\bibitem{Io} A. Iordanova,
RHIC\& AGS Annual User's Meeting: Workshop
  on U+U and Cu+Au shape studies (2014).

\bibitem{HHH12} Y. Hirono, M. Hongo, and T. Hirano,
Phys. Rev.  {\bf{C90}}, 021903  (2014).

\bibitem{Ca11}  W. Cassing, O. Linnyk, T. Steinert, and V. Ozvenchuk, Phys. Rev. Lett. {\bf{110}}, 182301  (2013);
T. Steinert and W. Cassing,  Phys. Rev. {\bf C 89}, 035203 (2014).

\bibitem{RS14} A. Rybicki and A. Szczurek,
arXiv:1405.6860;
Phys. Rev. {\bf C87}, 054909 (2013).
  Phys. Rev. {\bf C75}, 054903 (2007).

\bibitem{GKR14}U. G\"ursoy, D. Kharzeev and K. Rajagopal,
Phys. Rev. {\bf C89}, 054905 (2014).

\bibitem{VT11} V. Voronyuk, V.D Toneev, W. Cassing, E.L.
  Bratkovskaya, V.P. Konchakovski, and S.A. Voloshin,
Phys. Rev.   {\bf C83}, 054911 (2011).

\bibitem{WMZS14} J. Wang, Y.G. Ma, G.Q. Zhang, and W.Q. Shen,
arXiv:1411.1812.

\bibitem{JCG04} S. Juchem, W. Cassing, and C. Greiner, Phys. Rev. {\bf
  D69}, 025006 (2004); Nucl. Phys. {\bf A743}, 92 (2004).

\bibitem{CB09} W. Cassing, E. L. Bratkovskaya, Nucl. Phys. {\bf A831},
  215 (2009); Phys. Rev. {\bf C78}, 034919 (2008); W. Cassing,
  Nucl. Phys. {\bf A791}, 365 (2007).

\bibitem{Ca09} W. Cassing, E. Phys. J. ST {\bf 168}, 3 (2009).

\bibitem{Ko12} V. P. Konchakovski, E. L. Bratkovskaya, W. Cassing, V. D. Toneev, and V. Voronyuk, Phys. Rev. {\bf C85}, 011902 (2012).

\bibitem{Li11} O. Linnyk, E.L. Bratkovskaya, V. Ozvenchuk, W. Cassing, and C.M. Ko, Phys. Rev. {\bf C84}, 054917 (2011);
  O. Linnyk, W. Cassing, J. Manninen, E.L. Bratkovskaya, and C.M. Ko, Phys. Rev. {\bf C85}, 024910 (2012); O. Linnyk, W. Cassing, J. Manninen, E.L. Bratkovskaya, 
  P.B. Gossiaux, J. Aichelin, T. Song, and C.M. Ko, Phys. Rev. {\bf C87}, 014905
  (2013); O. Linnyk, V.P. Konchakovski, W. Cassing, and E.L. Bratkovskaya, Phys. Rev. {\bf C88},  034904 (2013).

\bibitem{To12} V. D. Toneev et al., Phys. Rev. {\bf C85}, 034910 (2012).

\bibitem{EC96} W. Ehehalt and W. Cassing, Nucl. Phys. {\bf A602}, 449 (1996).

\bibitem{CB99} W. Cassing and E. L. Bratkovskaya, Phys. Rep. {\bf 308},
  65 (1999).

\bibitem{CBJ00} W. Cassing, E. L. Bratkovskaya and S. Juchem,
  Nucl. Phys. {\bf A674}, 249 (2000).


\bibitem{BS12} A. Bzdak and V. Skokov,
  Phys. Lett. {\bf B710}, 171 (2012).

\bibitem{DH12} W.-T. Deng and X.-G. Huang, Phys. Rev. {\bf C85}, 044907 (2012).

\bibitem{Tu13}K. Tuchin,
Int. J. Mod. Phys. {\bf E23}, No. 1,  1430001 (2014).

\bibitem{KBC12} V. P. Konchakovski, E. L. Bratkovskaya, W. Cassing,
  V. D. Toneev, S. A. Voloshin, and V. Voronyuk,
Phys. Rev. {\bf C85}, 044922 (2012).

\bibitem{BBM11} P. Bozek, W. Broniowski and J. Moreira,
Phys. Rev.  {\bf     C83},  034911 (2011).

\bibitem{BLO11} R. S. Bhalerao, M. Luzum and J.-Y. Ollitrault,
Phys. Rev. {\bf C84},  054901 (2011).

\bibitem{Elena14} E. L. Bratkovskaya,
 arXiv:1408.3674; O. Linnyk, W. Cassing, and E.L. Bratkovskaya, Phys. Rev. {\bf C89},  034908 (2014).


\end{thebibliography}
\end{document}